\documentclass[aps,twocolumn]{revtex4}

\usepackage{amsmath}

\usepackage{amsmath}
\usepackage{graphicx}

\usepackage{amsmath}
\usepackage{stmaryrd}

\usepackage{epsfig}

\def\ov#1{\overline{#1}}

\newcommand{\bc}{\begin{center}}
\newcommand{\ec}{\end{center}}
\newcommand{\bt}{\begin{tabbing}}
\newcommand{\et}{\end{tabbing}} 
\newcommand{\be}{\begin{eqnarray*}}
\newcommand{\ee}{\end{eqnarray*}}

\begin{document}

\title{Notes on the Weierstrass Elliptic Function}

\author{Alain J. Brizard}
\affiliation{Department of Physics, Saint Michael's College, Colchester, VT 05439, USA}

\begin{abstract}
A consistent notation for the Weierstrass elliptic function $\wp(z;g_{2},g_{3})$, for $g_{2} > 0$ and arbitrary values of $g_{3}$ and $\Delta \equiv g_{2}^{3} - 27 g_{3}^{2}$, is introduced based on the parametric solution for the motion of a particle in a cubic potential. These notes provide a roadmap for the use of {\sf Mathematica} to calculate the half-periods $(\omega_{1},\omega_{3},\omega_{2} \equiv \omega_{1} + \omega_{3})$ of the Weierstrass elliptic function.
\end{abstract}

\begin{flushright}
October 13, 2015
\end{flushright}


\maketitle

\section{Introduction}

The present paper introduces a consistent notation for the Weierstrass elliptic function $\wp(z; g_{2}, g_{3})$, the roots $({\sf e}_{1},{\sf e}_{2},
{\sf e}_{3})$, and the half-periods $(\omega_{1},\omega_{3},\omega_{2} \equiv \omega_{1} + \omega_{3})$, where each half-period $\omega_{k}(g_{2},g_{3})$ and its associated root ${\sf e}_{k}$ satisfy the definition $\wp(\omega_{k}; g_{2}, g_{3}) \equiv {\sf e}_{k}$. The notation for 
the half-periods $(\omega_{1},\omega_{2},\omega_{3})$ depends on the signs of $g_{3}$ and the modular discriminant defined as $\Delta \equiv 
g_{2}^{3} - 27\,g_{3}^{2}$. 

The standard notation for the Weierstrass half-periods $(\ov{\omega}_{1},\ov{\omega}_{2},\ov{\omega}_{3})$ when $(g_{3},\Delta) = (\pm,+)$ follows Refs.~\cite{NIST_Weierstrass, Lawden}. In the  case $(g_{3},\Delta) = (+,+)$, the half-period $\ov{\omega}_{1}(g_{2},g_{3}) = \omega$ is real and the half-period $\ov{\omega}_{3}(g_{2},g_{3}) = \omega^{\prime} = i\,|\omega^{\prime}|$ is imaginary. The half-periods $(\ov{\omega}_{1},\ov{\omega}_{3},\ov{\omega}_{2} \equiv \ov{\omega}_{1} + \ov{\omega}_{3})$, therefore, form a rectangular lattice with the fundamental period parallelogram defined with corners located at $(0,2\omega, 2\omega^{\prime}, 2\omega + 2\omega^{\prime})$. When $g_{3}$ changes sign, i.e., 
$(g_{3},\Delta) = (-,+)$, the fundamental period parallelogram is rotated clockwise by $90^{o}$ to a new parallelogram with corners located at  $(0,-2i\,\omega, 2|\omega^{\prime}|, 
2|\omega^{\prime}| - 2i\,\omega)$, where we used $-i\omega^{\prime} = |\omega^{\prime}|$. The notation for the half-periods $(\omega_{1},\omega_{3}, \omega_{2} \equiv \omega_{1} + \omega_{3})$ adopted here will also follow this standard notation when $(g_{3},\Delta) = (\pm,+)$.

When $\Delta$ changes sign, i.e., $(g_{3},\Delta) = (+,-)$, the standard notation \cite{NIST_Weierstrass} states that the half-period $\ov{\omega}_{1}(g_{2},g_{3}) = \Omega$ is real and the half-period $\ov{\omega}_{3}(g_{2},g_{3}) = \Omega/2 + \Omega^{\prime}$ is complex-valued (where $\Omega^{\prime}$ is purely imaginary). The half-periods $(\ov{\omega}_{1},\ov{\omega}_{3},\ov{\omega}_{2} \equiv \ov{\omega}_{1} - \ov{\omega}_{3})$ form a rhombic lattice with the fundamental period parallelogram defined with corners located at $(0,2\Omega, \Omega + 2\Omega^{\prime}, \Omega - 2\Omega^{\prime})$. When $g_{3}$ changes sign, i.e., $(g_{3},\Delta) = (-,-)$, the fundamental period parallelogram is rotated clockwise by $90^{o}$ to the new fundamental period parallelogram with corners located at $(0,-2i\,\Omega, 2|\Omega^{\prime}| - i\,\Omega, -2|\Omega^{\prime}| - i\,\Omega)$, i.e., $\ov{\omega}_{2}$ is now defined as $\ov{\omega}_{2} = \ov{\omega}_{3} - \ov{\omega}_{1}$ when $(g_{3},\Delta) = (-,-)$. The new notation for the half-periods $(\omega_{1},\omega_{2},\omega_{3})$ adopted here when $(g_{3},\Delta) = (\pm,-)$ will instead assume that 
$\omega_{2} \equiv \omega_{1} + \omega_{3}$, as in the standard case $(g_{3},\Delta) = (\pm,+)$.

The purpose of the present notes is to introduce a consistent classification of the roots and half-periods based on the orbit solutions derived for all energy levels associated with unbounded and bounded motion in a cubic potential. Additional notes on the analysis of periodic solutions in classical mechanics involving doubly-periodic elliptic functions are presented in Refs.~\cite{Brizard_2007,Brizard_2009,Brizard_Lag}.

The remainder of this paper is organized as follows. In Sec.~\ref{sec:cubic}, we introduce a parameterization of the turning points $x_{k}(\phi)$ for particle motion in a cubic potential based on the definition $g_{3}(\phi) = \cos\phi \equiv -\,2E$, where $g_{2} = 3$ and $E$ denotes the particle's total energy. In Sec.~\ref{sec:Weierstrass}, we introduce a consistent formulation of the Weierstrass elliptic function $\wp(z;g_{2},g_{3})$ associated with the parameterization introduced in Sec.~\ref{sec:cubic}. In Sec.~\ref{sec:W_orbits}, the Weierstrass orbit solutions \eqref{eq:x_wp} are presented for all energy-level classes for the motion in a cubic potential. Lastly, the Jacobi orbit solutions are presented in Sec.~\ref{sec:J_orbits} based on connection relations between the Weierstrass and Jacobi elliptic functions.

\section{\label{sec:cubic}Motion in a Cubic Potential}

We consider the motion of a (unit-mass) particle in the cubic potential
\begin{equation}
V(x) \;=\; \frac{3}{2}\,x \;-\; 2\,x^{3},
\label{eq:V_def}
\end{equation}
which has an inflection point at $x = 0$, a minimum at $x = -1/2$, with $V(-1/2) = -1/2$ and $V^{\prime\prime}(-1/2) = 6$ (i.e., $x = -1/2$ is a stable equilibrium point), and a maximum at $x = 1/2$, with $V(1/2) = 1/2$ and $V^{\prime\prime}(1/2) = -\,6$ (i.e., $x = 1/2$ is an unstable equilibrium point). The period of small oscillations near the stable equilibrium point at $x = -1/2$ is $2\pi/\sqrt{6}$.

\begin{figure}
\epsfysize=2in
\epsfbox{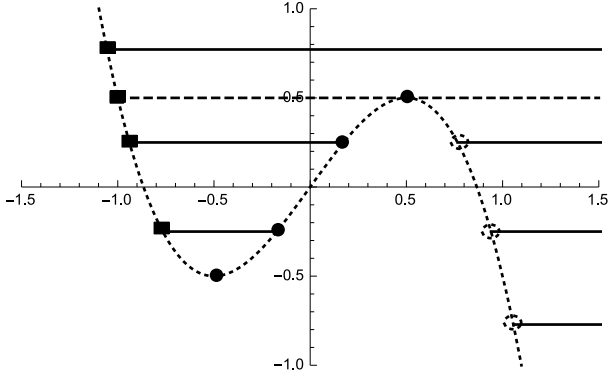}
\caption{Energy levels for the particle orbits in the cubic potential \eqref{eq:V_def}, shown here as a dotted curve. Solid squares represent the locations of the root $x_{3}(\phi)$, solid circles represent the locations of the root $x_{2}(\phi)$, and open circles represent the locations of the root $x_{1}(\phi)$. The energy level $(E = 1/2)$ labeled by a dashed line denotes the bounded and unbounded separatrix orbits.}
\label{fig:Orbits}
\end{figure}

The equation of motion for a particle of unit mass and total energy $E$ is
\begin{equation}
\dot{x}^{2} \;=\; 2\,E \;-\; 3\,x \;+\; 4\,x^{3} \;\equiv\; 4\,x^{3} \;-\; g_{2}\,x \;-\; g_{3},
\label{eq:W_form}
\end{equation}
where $\dot{x} = dx/dt$ denotes the particle velocity. The right side of Eq.~\eqref{eq:W_form} is expressed in the Weierstrass standard form with the invariants $g_{2} = 3$ and $g_{3} = -\,2E$, where bounded and unbounded orbit solutions are classified according to four energy levels summarized in Table \ref{tab:energy_levels}. The total energy $E \equiv -\frac{1}{2}\,\cos\phi$ is defined in terms of the Weierstrass phase $\phi$, which is defined along a continuous path in the complex-$\phi$ plane:
\begin{equation}
\phi \;=\; \left\{ \begin{array}{lr}
-i\,\psi & (\psi \geq 0) \\
\varphi & (0 \leq \varphi \leq \pi/2) \\
\pi - \varphi & (0 \leq \varphi \leq \pi/2) \\
\pi + i\,\psi & (\psi \geq 0)
\end{array} \right.
\label{eq:W_phi}
\end{equation}
which allows us to go continuously from $E < -1/2$ (lowest energy level in Fig.~\ref{fig:Orbits}) to $E > 1/2$ (highest energy level in Fig.~\ref{fig:Orbits}).

\begin{table}
\caption{\label{tab:energy_levels}Energy levels $E = -\frac{1}{2}\cos\phi$ defined in terms of the Weierstrass phase \eqref{eq:W_phi}.} 
\begin{ruledtabular} 
\begin{tabular}{lcc}
Energy Levels                             & $\phi$                 & $E = -\frac{1}{2}\cos\phi$  \\ \hline 
(I)                & $-i\,\psi$             & $E = -\frac{1}{2}\cosh\psi \leq -\frac{1}{2}$ \\
(II)               & $\varphi$             & $-\frac{1}{2} \leq E = -\frac{1}{2}\cos\varphi \leq 0$  \\ 
(III)              & $\pi - \varphi$     & $0 \leq E = \frac{1}{2}\cos\varphi \leq \frac{1}{2}$ \\
(IV)             & $\pi + i\,\psi$       & $E = \frac{1}{2}\cosh\psi \geq \frac{1}{2}$
\end{tabular}
\end{ruledtabular}
\end{table}

\begin{figure}
\epsfysize=2in
\epsfbox{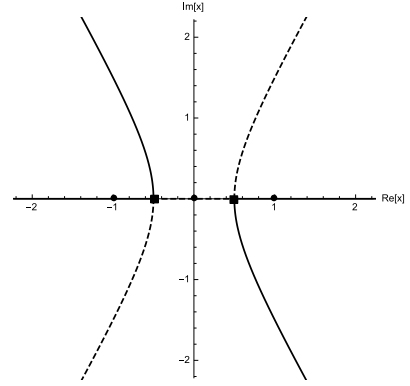}
\caption{Graphs of the roots $(x_{1},x_{2},x_{3})$ defined in Eq.~\eqref{eq:roots} and shown in Table \ref{tab:roots}. The real and imaginary parts of $x_{1}(\phi)$ and $x_{3}(\phi)$ are shown as solid curves while the real and imaginary parts of $x_{2}(\phi)$ are shown as dashed curves. In region (I), $x_{1} > 1$ is real and $x_{2} = x_{3}^{*} = a - i b$, with $a < -1/2$. In regions (II)-(III), all roots are real, with $-1 < x_{3} < x_{2} < x_{1} < 1$, and $x_{2}$ and $g_{3}$ have opposite signs. In region (IV), $x_{3} < -1$ is real and $x_{1} = x_{2}^{*} = a - i b$, with $a > 1/2$. The symmetry properties \eqref{eq:symmetry_roots} are clearly apparent with $x_{2}(\phi)$ and ${\rm Im}(x_{1}) = -\,{\rm Im}(x_{3}) \leq 0$.}
\label{fig:Weierstrass_roots}
\end{figure}

The turning points $(x_{1},x_{2},x_{3})$, where the velocity $\dot{x}$ vanishes in Eq.~\eqref{eq:W_form}, are defined as the roots of the energy equation $E  = -\frac{1}{2}\,\cos\phi = V(x)$:
\begin{equation}
\left. \begin{array}{rcl}
x_{1}(\phi) & = & \cos(\phi/3) \\
x_{2}(\phi) & = & - \cos[(\pi + \phi)/3] \\
x_{3}(\phi) & = & - \cos[(\pi - \phi)/3]
\end{array} \right\},
\label{eq:roots}
\end{equation}
which satisfy the condition $x_{1} + x_{2} + x_{3} = 0$. The turning points \eqref{eq:roots} are shown in Table \ref{tab:roots} (as well as 
Figs.~\ref{fig:Orbits} and \ref{fig:Weierstrass_roots}) for each energy region defined in Table \ref{tab:energy_levels}. The equation of motion \eqref{eq:W_form} can, therefore, also be expressed as
\begin{equation}
\dot{x}^{2} \;\equiv\; 4\,(x - x_{1})\,(x - x_{2})\,(x - x_{3}),
\label{eq:W_form_roots}
\end{equation}
with the roots defined in Eq.~\eqref{eq:roots}.

\begin{table}
\caption{\label{tab:roots}Turning points $x_{k}(\phi)$ defined in Eq.~\eqref{eq:roots}.} 
\begin{ruledtabular} 
\begin{tabular}{lccc}
Energy    & $x_{1}(\phi)$                   & $x_{2}(\phi)$                       &  $x_{3}(\phi)$                          \\ \hline 
(I)            & $\cosh(\psi/3)$                 & $-\cos[(\pi - i\psi)/3]$         &  $-\cos[(\pi + i\psi)/3]$     \\
(II)          & $\cos(\varphi/3)$              & $-\cos[(\pi + \varphi)/3]$     &  $-\cos[(\pi - \varphi)/3]$  \\ 
(III)         & $\cos[(\pi - \varphi)/3]$   & $\cos[(\pi + \varphi)/3]$       &  $-\cos(\varphi/3)$ \\
(IV)         & $\cos[(\pi + i\psi)/3]$       & $\cos[(\pi - i\psi)/3]$           &  $- \cosh(\psi/3)$
\end{tabular}
\end{ruledtabular}
\end{table}

The orbit solutions for Eq.~\eqref{eq:W_form} are expressed in terms of the Weierstrass elliptic function 
\begin{equation}
x(t;x_{0}) \;=\; \wp(t + \gamma; 3, \cos\phi),
\label{eq:x_wp}
\end{equation}
where the constant $\gamma(g_{2}, g_{3}; \Delta)$ is determined from the initial condition 
\begin{equation}
x_{0}(\phi) \;=\;  \wp(\gamma; 3, \cos\phi).
\label{eq:x0_sol}
\end{equation}
When the motion is periodic, the period $2\omega_{k}(g_{2},g_{3})$ depends on $g_{2} = 3$, $g_{3} = \cos\phi$, and the sign of the modular discriminant $\Delta \equiv g_{2}^{3} - 27\,g_{3}^{2}$: 
\[ \Delta \;=\; 27\,\sin^{2}\phi \;=\; \left\{ \begin{array}{lr}
-27\,\sinh^{2}\psi \leq 0 & (|E| \geq 1/2) \\
 & \\
 27\,\sin^{2}\varphi \geq 0 & (|E| \leq 1/2)
 \end{array} \right. \]
 Here, the half-periods $\omega_{k} = (\omega_{1},\omega_{2} \equiv \omega_{1} + \omega_{3}, \omega_{3})$ satisfy the definition 
 \begin{equation}
 \wp(\omega_{k}; 3,\cos\phi) \;\equiv\; x_{k}(\phi).
 \label{eq:wp_x}
 \end{equation}
Figure \ref{fig:Weierstrass_roots} shows the roots \eqref{eq:roots}, which are also shown in Table \ref{tab:roots}. In region (I), where $E < -1/2$, we note that $x_{1}^{I} >1$ and $x_{2}^{I} = x_{3}^{I*}$. In regions (II)-(III), where $-1/2 < E < 1/2$, the three turning points are real, with $x_{3} < x_{2} < x_{1}$. In region (IV), where $E > 1/2$, we note that 
$x_{3}^{IV} < -1$ and $x_{1}^{IV} =  x_{2}^{IV*}$. Moreover, in Table \ref{tab:roots}, we observe the following symmetry properties (see
Fig.~\ref{fig:Weierstrass_roots})
\begin{equation}
\left. \begin{array}{rcl}
x_{2}^{I} & = & - x_{2}^{IV}\;\;\;{\rm and}\;\;\;x_{2}^{II} =  -x_{2}^{III} \\
 &  & \\
x_{1}^{I} & = & - x_{3}^{IV}\;\;\;{\rm and}\;\;\;x_{1}^{II} =  -x_{3}^{III}
\end{array} \right\}. 
\label{eq:symmetry_roots}
\end{equation}
Lastly, we note that bounded periodic motion exists between the turning points  $x_{3} < x_{2}$ only for the energy levels $|E| < 1/2$.

\section{\label{sec:Weierstrass}Weierstrass Elliptic Function}

Before finding explicit expressions for the solutions \eqref{eq:x_wp}, we need to explore the properties of the Weierstrass elliptic function 
$\wp(z; g_{2}, g_{3})$, with the invariants $(g_{2},g_{3}; \Delta)$ defined as
\begin{equation}
\left. \begin{array}{rcl}
g_{2} & = & 3\;\beta^{2} \\
g_{3} & = & \beta^{3}\;\cos\phi \\
\Delta & = & g_{2}^{3} - 27\,g_{3}^{2} \;=\; 27\,\beta^{6}\,\sin^{2}\phi
\end{array} \right\}, 
\label{eq:g3_beta}
\end{equation}
where 
$\beta$ is an arbitrary (real or complex) parameter and the phase $\phi$ may be real or complex. The differential equation for the Weierstrass elliptic function $w \equiv \wp(z; g_{2}, g_{3})$ is
\begin{equation}
\left(\frac{dw}{dz}\right)^{2} \;=\; 4\,w^{3} \;-\; g_{2}\,w \;-\; g_{3} \;\equiv\; P(w),
\label{eq:wp_diffeq}
\end{equation}
where the cubic polynomial $P(w)$ has three roots $({\sf e}_{1}, {\sf e}_{2}, {\sf e}_{3})$ defined as
\begin{equation}
\left. \begin{array}{rcl}
{\sf e}_{1}(\phi) & = & \beta\,\cos(\phi/3) \\
{\sf e}_{2}(\phi) & = & -\,\beta \cos[(\pi + \phi)/3] \\
{\sf e}_{3}(\phi) & = & -\,\beta \cos[(\pi - \phi)/3]
\end{array} \right\},
\label{eq:roots_e}
\end{equation}
where we used the definitions \eqref{eq:g3_beta}. The roots \eqref{eq:roots} are, of course, obtained from Eq.~\eqref{eq:roots_e} through the identity $x_{k}(\phi) \equiv 
{\sf e}_{k}(\phi)/\beta$. 

The polynomial $P(w) = 4\,w^{3} - g_{2}\,w - g_{3}$ has a minimum at $w = \sqrt{g_{2}/12} \equiv w_{0}$ and a maximum at $w = -\sqrt{g_{2}/12} = -\,w_{0}$, where 
$P(\pm w_{0}) = \mp\sqrt{g_{2}^{3}/27} - g_{3}$. We note that the extrema $\pm w_{0}$ are real only if $g_{2} > 0$ and $g_{3}$ is real. Hence, using the discriminant $\Delta \equiv g_{2}^{3} - 27\,g_{3}^{2}$, the maximum $P(-w_{0}) = \sqrt{g_{3}^{2} + \Delta/27} - g_{3} > 0$ if $\Delta > 0$, and the three roots \eqref{eq:roots_e} are real when $g_{3} > 0$. When $g_{3} < 0$, on the other hand, the minimum $P(w_{0}) = -\sqrt{g_{3}^{2} + \Delta/27} + |g_{3}| < 0$ if $\Delta > 0$ and the three roots \eqref{eq:roots_e} are once again real. Lastly, only one root is real when $\Delta < 0$: ${\sf e}_{1}$ is real when $g_{3} > 0$ (with ${\sf e}_{2} = {\sf e}_{3}^{*}$) or ${\sf e}_{3}$ is real when $g_{3} < 0$ (with 
${\sf e}_{1} = {\sf e}_{2}^{*}$).

\subsection{Inversion formulas}

Equation \eqref{eq:wp_diffeq} can be transformed as
\begin{eqnarray*}
\lambda^{6}\;\left(\frac{dw}{dz}\right)^{2} & = & \left[ \frac{d(\lambda^{2}w)}{d(z/\lambda)}\right]^{2} \\
 & = & 4\,\left(\lambda^{2}w\right)^{3} \;-\; \left(\lambda^{4}\,g_{2}\right)\,\left(\lambda^{2}w\right) \;-\; \left(\lambda^{6}\,g_{3}\right) \\
 & \equiv & \left(\frac{d\ov{w}}{d\tau}\right)^{2} \;=\; 4\,\ov{w}^{3} \;-\; \ov{g}_{2}\;\ov{w} \;-\; \ov{g}_{3},
\end{eqnarray*}
where $\ov{w} \equiv \lambda^{2}\,w$ and $\tau \equiv z/\lambda$. Since $\ov{w}$ has the Weierstrass solution
\begin{eqnarray*}
\ov{w}(\tau) \;=\; \wp(\tau;\; \ov{g}_{2}, \ov{g}_{3}) & = & \wp\left(\lambda^{-1}z;\frac{}{} \lambda^{4}\,g_{2},\; \lambda^{6}\,g_{3}\right) \nonumber \\
 & \equiv & \lambda^{2}\;\wp(z;\;g_{2}, g_{3}),
\end{eqnarray*}
we obtain the identity
\begin{equation}
\wp(z;\;g_{2}, g_{3}) \;\equiv\; \lambda^{-2}\wp\left(\lambda^{-1}z;\frac{}{} \lambda^{4}\,g_{2},\; \lambda^{6}\,g_{3}\right).
\label{eq:lambda_sol}
\end{equation}
Hence, if $\lambda = -1$, the Weierstrass elliptic function is shown to have even parity: 
\begin{equation}
\wp(-z; g_{2},g_{3}) \;=\; \wp(z; g_{2},g_{3}). 
\label{eq:even}
\end{equation}
Next, if $\lambda = -\,i$, we find the $g_{3}$-inversion formula
\begin{equation}
\wp\left(z;\frac{}{}g_{2},g_{3}\right) \;\equiv\; -\;\wp(i\,z;\;g_{2}, |g_{3}|).
\label{eq:g3_inv}
\end{equation}
The $g_{2}$-inversion formula, on the other hand, is obtained with $\lambda = \exp(-i\pi/4)$ so that
\begin{equation}
\wp(z;\;g_{2}, g_{3}) \;\equiv\; i\,\wp\left(e^{i\pi/4}\,z;\frac{}{}|g_{2}|, i\,g_{3}\right).
\label{eq:g2_inv}
\end{equation}
Lastly, by substituting $\lambda \equiv \beta^{\frac{1}{2}}$ into Eq.~\eqref{eq:lambda_sol}, we find
\begin{equation}
\wp(t + \gamma;\;3, \cos\phi) \;\equiv\; \beta^{-1}\wp\left(\beta^{-\frac{1}{2}}(t + \gamma);\frac{}{} g_{2}, g_{3}\right),
\label{eq:beta_sol}
\end{equation}
where $(g_{2}, g_{3})$ are given by Eq.~\eqref{eq:g3_beta}. Hence, the general orbit solution \eqref{eq:x_wp} can be mapped into the general solution 
$\wp(z; g_{2}, g_{3})$ for the Weierstrass differential equation \eqref{eq:wp_diffeq}, with $z \equiv \beta^{-\frac{1}{2}}(t + \gamma)$.

\subsection{Weierstrass Half-periods}

We now explore the half-periods $\omega_{1}(g_{2},g_{3})$ and $\omega_{3}(g_{2},g_{3})$ for $g_{2} > 0$ and the four sign-pairings for 
$(g_{3},\Delta)$ shown in Table \ref{tab:periods}. Here, the half-periods $\omega$ and $\Omega$ are real, and the half-periods 
$\omega^{\prime} = i|\omega^{\prime}|$ and $\Omega^{\prime} = i|\Omega^{\prime}|$ are imaginary. When $g_{2} = 3\beta^{2}$ and $g_{3} =
\beta^{3} > 0$, so that $\Delta = 0$, we find
\begin{equation}
\omega\left(3\beta^{2}, \beta^{3}\right) \;=\; \Omega\left(3\beta^{2}, \beta^{3}\right) \;=\; \frac{\pi}{\sqrt{6\beta}} \;\equiv\; \Omega_{0}
\label{eq:Omega0_def}
\end{equation} 
and $\omega^{\prime} = \Omega^{\prime} = i\infty$. When $g_{2} = 3\beta^{2}$ and $g_{3} = 0$, we find 
\begin{equation}
\omega\left(3\beta^{2},0\right) \;=\; \frac{{\sf K}(1/2)}{(3\beta^{2})^{1/4}} \;\equiv\; \omega_{0}, 
\label{eq:omega0_def}
\end{equation}
where ${\sf K}(m)$ denotes the complete elliptic integral of the first kind. 

\begin{figure}
\epsfysize=1.8in
\epsfbox{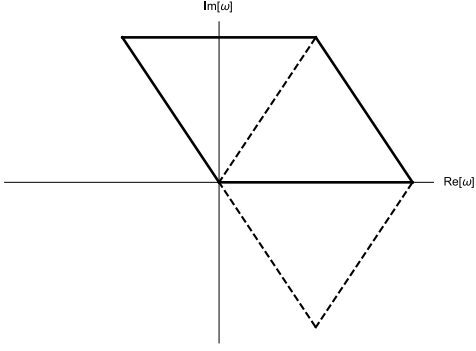}
\caption{New fundamental period parallelogram \eqref{eq:new_FPP}, shown as a solid rhombus with corners located at $(0, 2\,\Omega, -\Omega + 2\,\Omega^{\prime}, \Omega + 2\,\Omega^{\prime})$, and standard fundamental period parallelogram \eqref{eq:standard_FPP}, shown as a dashed rhombus with corners located at $(0, 2\,\Omega, \Omega + 2\,\Omega^{\prime}, \Omega - 2\,\Omega^{\prime})$, for the case $(g_{3},\Delta) = 
(+,-)$. Here, $\Omega$ is purely real and $\Omega^{\prime}$ is purely imaginary.}
\label{fig:FPP}
\end{figure}

Figure \ref{fig:FPP} compares the new fundamental period parallelogram (shown as a solid rhombus) for the case $(g_{3},\Delta) = (+,-)$ with the standard fundamental period parallelogram (shown as a dashed rhombus) as defined in Ref.~\cite{NIST_Weierstrass}. The corners of the new rhombus are located at 
\begin{equation}
(0,2\,\omega_{1},2\,\omega_{3},2\,\omega_{2}) = (0, 2\,\Omega, -\Omega + 2\,\Omega^{\prime}, \Omega + 2\,\Omega^{\prime}),
\label{eq:new_FPP}
\end{equation}
where $\omega_{2} \equiv \omega_{1} + \omega_{3}$, while the corners of the standard rhombus are located at
 \begin{equation}
(0,2\,\ov{\omega}_{1},2\,\ov{\omega}_{3},2\,\ov{\omega}_{2}) = (0, 2\,\Omega, \Omega + 2\,\Omega^{\prime}, \Omega - 2\,\Omega^{\prime}), 
\label{eq:standard_FPP}
\end{equation}
where $\ov{\omega}_{2} \equiv \ov{\omega}_{1} - \ov{\omega}_{3}$. Hence, we immediately see that both fundamental period parallelograms identify $\omega_{1} = \Omega = \ov{\omega}_{1}$ as real, while the new fundamental period parallelogram identifies the new half-period $\omega_{2}$ as the standard half-period $\ov{\omega}_{3}$ in the standard fundamental period parallelogram. We note that the standard half-periods $(\ov{\omega}_{2},\ov{\omega}_{3})$ satisfy $\ov{\omega}_{2} \equiv \ov{\omega}_{1} - \ov{\omega}_{3} = \ov{\omega}_{3}^{*}$, while the new half-periods $(\omega_{2},\omega_{3})$ satisfy $\omega_{2} \equiv \omega_{1} + \omega_{3} \;=\; -\,\omega_{3}^{*}$.
Both definitions satisfy the relations ${\sf e}_{2} = \wp(\ov{\omega}_{2}; g_{2}, g_{3}) = [\wp(\ov{\omega}_{3}; g_{2}, g_{3})]^{*} = {\sf e}_{3}^{*}$ and ${\sf e}_{2} = 
\wp(\omega_{2}; g_{2}, g_{3}) = [\wp(-\omega_{3}; g_{2}, g_{3})]^{*} = {\sf e}_{3}^{*}$, which follows from the fact that the Weierstrass elliptic function has even parity, i.e., 
$\wp(-\omega_{3}) = \wp(\omega_{3})$. The advantage of the new definition of $\omega_{2}$ is that it is consistently defined as $\omega_{2} \equiv \omega_{1} + \omega_{3}$ 
for all values of $(g_{3},\Delta)$.

\begin{table}
\caption{\label{tab:periods}Half-periods $\omega_{1}(g_{2},g_{3})$ and $\omega_{3}(g_{2},g_{3})$ for $g_{2} = 3\,\beta^{2}$, $g_{3} = \beta^{3}
\cos\phi$, and $\Delta = 27\beta^{6}\sin^{2}\phi$, with $\omega_{2} \equiv \omega_{1} + \omega_{3}$.} 
\begin{ruledtabular} 
\begin{tabular}{ccccc}
 & $(g_{3},\Delta)$  & $\omega_{1}$                                    &  $\omega_{3}$                                 &  $\omega_{2}$                      \\ \hline 
(I) & $(+,-)$                & $\Omega$                                          &  $-\Omega/2 + \Omega^{\prime}$   &  $\Omega/2 + \Omega^{\prime}$  \\
 & $(+,0)$               & $\Omega_{0}$                                   &  $i\,\infty$                                        & $i\,\infty$  \\       
(II) & $(+,+)$               & $\omega$                                           &  $\omega^{\prime}$                         &  $\omega + \omega^{\prime}$    \\ 
 & $(0,+)$               & $\omega_{0}$                                    & $i\,\omega_{0}$                                 &  $(1 + i)\,\omega_{0}$ \\
(III) & $(-,+)$                & $-i\,\omega^{\prime} = |\omega^{\prime}|$          &  $-i\,\omega$                    & $|\omega^{\prime}| - i\omega$  \\
 & $(-,0)$               & $\infty$                                                & $-i\,\Omega_{0}$                               & $\infty$ \\
(IV) & $(-,-)$                & $|\Omega^{\prime}| + i\,\Omega/2$      &  $-i\,\Omega$                                    & $|\Omega^{\prime}| - i\,\Omega/2$
\end{tabular}
\end{ruledtabular}
\end{table}

\begin{figure}
\epsfysize=4in
\epsfbox{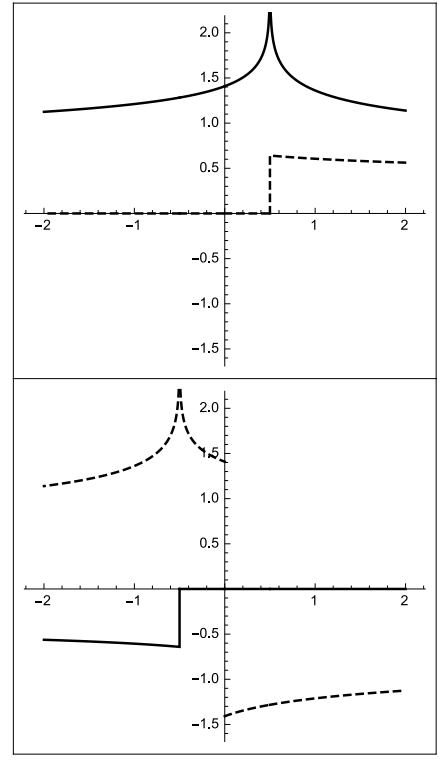}
\caption{Plots of $\omega_{1}$ (top; real part = solid and imaginary part = dashed), Eq.~\eqref{eq:omega1_int}, and $\omega_{3}$ (bottom; real part = solid and imaginary part = dashed), Eq.~\eqref{eq:omega3_int}, as functions of energy $-2 \leq E = -\,g_{3}/2 \leq 2$, with $\beta = 1$ in Eqs.~\eqref{eq:g3_beta} and \eqref{eq:roots_e}. When $g_{3}$ changes sign, $\omega_{1}^{+} = \Omega\;(E \leq -1/2)$ or $= \omega\;(-1/2 \leq E \leq 0)$ transforms into $\omega_{3}^{-} = -i\,\omega_{1}^{+} = -i\,\omega\;(0 \leq E \leq 1/2)$ or $= -i\,\Omega\;(E \geq 1/2)$, while $\omega_{3}^{+} = -\,\Omega/2 + \Omega^{\prime}\;(E \leq -1/2)$ or $= \omega^{\prime}\;(-1/2 \leq E \leq 0)$ transforms into $\omega_{1}^{-} = -i\,\omega_{3}^{+} =  |\omega^{\prime}|\;(0 \leq E \leq 1/2)$ or $= i\,\Omega/2 + |\Omega^{\prime}|\;(E \geq 1/2)$.}
\label{fig:Weierstrass_half-periods}
\end{figure}

The definitions in Table \ref{tab:periods} for the half-periods satisfy Eq.~\eqref{eq:wp_x}, while the symmetry properties associated with the sign inversion for $g_{3}$ (for $g_{2} > 0$) are satisfied by Eq.~\eqref{eq:g3_inv} between $(g_{3},\Delta) = (\pm,-)$ in regions (I) and (IV):
\begin{eqnarray}
{\sf e}_{1}^{-} & = & \wp(|\Omega^{\prime}| + i\Omega/2; g_{2}, g_{3} < 0) \nonumber \\
 & = & -\;\wp(\Omega^{\prime} - \Omega/2; g_{2}, |g_{3}|) \;=\; -\,{\sf e}_{3}^{+}, \label{eq:x_13_IIV} \\
 {\sf e}_{2}^{-} & = & \wp(|\Omega^{\prime}| - i\Omega/2; g_{2}, g_{3} < 0) \nonumber \\
 & = & -\;\wp(\Omega^{\prime} + \Omega/2; g_{2},  |g_{3}|) \;=\; -\,{\sf e}_{2}^{+}, \label{eq:x_2_IIV} \\
 {\sf e}_{3}^{-} & = & \wp(-i\Omega; g_{2}, g_{3} < 0) \nonumber \\
 & = & -\;\wp(\Omega; g_{2},  |g_{3}|) \;=\; -\,{\sf e}_{1}^{+}, \label{eq:x_31_IIV}
\end{eqnarray}
and  $(g_{3},\Delta) = (\pm,+)$ in regions (II) and (III):
\begin{eqnarray}
{\sf e}_{1}^{-} & = & \wp(|\omega^{\prime}|; g_{2}, g_{3} < 0) \nonumber \\
 & = & -\;\wp(\omega^{\prime}; g_{2},  |g_{3}|) \;=\; -\,{\sf e}_{3}^{+}, \label{eq:x_13_IIII} \\
 {\sf e}_{2}^{-} & = & \wp(|\omega^{\prime}| - i\omega; g_{2}, g_{3} < 0) \nonumber \\
 & = & -\;\wp(\omega^{\prime} + \omega; g_{2},  |g_{3}|) \;=\; -\,{\sf e}_{2}^{+}, \label{eq:x_2_IIII} \\
 {\sf e}_{3}^{-} & = & \wp(-i\omega; g_{2}, g_{3} < 0) \nonumber \\
 & = & -\;\wp(\omega; g_{2},  |g_{3}|) \;=\; -\,{\sf e}_{1}^{+}. \label{eq:x_31_IIII}
\end{eqnarray}
From Eqs.~\eqref{eq:x_13_IIV}-\eqref{eq:x_31_IIV}, we immediately observe that $({\sf e}_{1}^{-} )^{*} = {\sf e}_{2}^{-}$ and $({\sf e}_{3}^{+} )^{*} = {\sf e}_{2}^{+}$, while 
${\sf e}_{3}^{-} = -\,{\sf e}_{1}^{+}$ are purely real.

Lastly, the direct calculations of the half-periods $\omega_{1}(g_{2},g_{3})$ and $\omega_{3}(g_{2},g_{3})$ are given in terms of the integrals \cite{NIST_Weierstrass}
\begin{eqnarray}
\omega_{1}^{\pm}(g_{2},g_{3}) & = & \int_{{\sf e}_{1}^{\pm}}^{\infty}\frac{dw}{\sqrt{P_{\pm}(w)}}, \label{eq:omega1_int} \\
\omega_{3}^{\pm}(g_{2},g_{3}) & = & \pm\,i\;\int_{-\infty}^{{\sf e}_{3}^{\pm}}\frac{dw}{\sqrt{|P_{\pm}(w)|}}, \label{eq:omega3_int}
\end{eqnarray}
where the sign $\pm$ in Eqs.~\eqref{eq:omega1_int}-\eqref{eq:omega3_int} is determined by the sign of $g_{3}$, $P_{\pm}(w) \equiv 4\,(w - {\sf e}_{1}^{\pm})(w - {\sf e}_{2}^{\pm})
(w - {\sf e}_{3}^{\pm})$, and the roots $({\sf e}_{1}^{\pm},{\sf e}_{2}^{\pm},{\sf e}_{3}^{\pm})$ are defined in Eq.~\eqref{eq:roots_e} in terms of the phase $\phi$. The half-periods
\eqref{eq:omega1_int}-\eqref{eq:omega3_int}, which are shown in Fig.~\ref{fig:Weierstrass_half-periods}, satisfy the $g_{3}$-inversion relations
\begin{eqnarray}
\omega_{3}^{-} & = & -i\int_{-\infty}^{{\sf e}_{3}^{-}}\frac{dw}{\sqrt{|P_{-}(w)|}} \;=\; -i\int^{-{\sf e}_{1}^{+}}_{-\infty}\frac{dw}{\sqrt{|P_{-}(w)|}} \nonumber \\
 & = & -i\int^{\infty}_{{\sf e}_{1}^{+}}\frac{dw}{\sqrt{P_{+}(w)}} \;=\; -i\,\omega_{1}^{+}, \label{eq:omega3_inv_int} \\
 \omega_{1}^{-} & = & \int^{\infty}_{{\sf e}_{1}^{-}}\frac{dw}{\sqrt{P_{-}(w)}} \;=\; \int_{-{\sf e}_{3}^{+}}^{\infty}\frac{dw}{\sqrt{P_{-}(w)}} \nonumber \\
 & = & \int_{-\infty}^{{\sf e}_{3}^{+}}\frac{dw}{\sqrt{|P_{+}(w)|}} \;=\; -i\,\omega_{3}^{+}, \label{eq:omega1_inv_int} 
\end{eqnarray}
where we used the relations \eqref{eq:x_13_IIV}-\eqref{eq:x_31_IIII}, with the identity $P_{-}(-w) \equiv -\,P_{+}(w)$. We note that $P_{+}(w) > 0$ in 
Eq.~\eqref{eq:omega3_inv_int} for $w > {\sf e}_{1}^{+}$, while $P_{+}(w) < 0$ in Eq.~\eqref{eq:omega1_inv_int} for $w < {\sf e}_{3}^{+}$, so that $P_{-}(-w) \equiv -\,P_{+}(w)
= |P_{+}(w)|$. The half-periods $\omega_{2}^{\pm} \equiv \omega_{1}^{\pm} + \omega_{3}^{\pm}$ are consistently defined in terms of Eqs.~\eqref{eq:omega1_int}-\eqref{eq:omega3_int}, with $\omega_{2}^{-} = -i\,\omega_{2}^{+}$.

\subsection{{\sf Mathematica} outputs}

\begin{figure}
\epsfysize=2in
\epsfbox{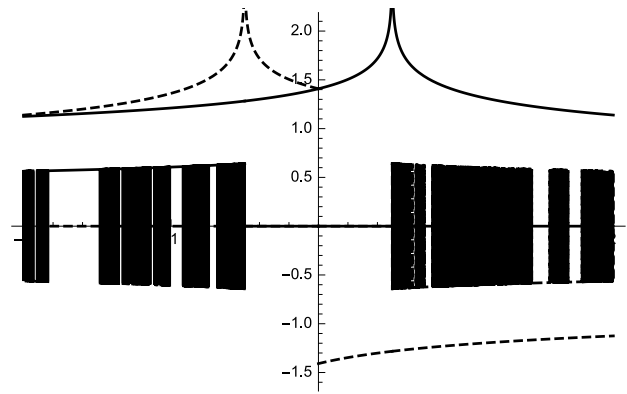}
\caption{Plots of {\sf Mathematica} real (solid) and imaginary (dashed) outputs $\{\omega_{a}, \omega_{b}\} =$ {\sf WeierstrassHalfPeriods}$(g_{2},g_{3})$.}
\label{fig:Weierstrass_rules}
\end{figure}

Table \ref{tab:periods_Mathematica} and Fig.~\ref{fig:Weierstrass_rules} show the outputs of the function {\sf WeierstrassHalfPeriods}$(g_{2},g_{3}) = \{\omega_{a},\omega_{b}\}$ from {\sf Mathematica}. Here, the first output $\omega_{a}$ is either $\omega_{1}^{+}$ if $E \leq 0$ or $\omega_{3}^{-} = -i\,\omega_{1}^{+}$ if $E > 0$. The second output $\omega_{b}$, on the other hand, is either $\omega_{2}^{\pm}$ if $|E| > 1/2$ or $\omega_{3}^{+}$ and $\omega_{1}^{-} = -i\,\omega_{3}^{+}$ if $|E| < 1/2$. The definitions of $(\omega_{1}^{\pm}, \omega_{3}^{\pm}, \omega_{2}^{\pm} \equiv \omega_{1}^{\pm} + \omega_{3}^{\pm})$ follow the notation presented in Table \ref{tab:periods}.

We note that, while {\sf Mathematica}'s output $\omega_{a}$ is well behaved for all values of $E$, the output $\omega_{b}$ is stable only when 
$|E| \leq 1/2$. When $|E| > 1/2$, however, the output $\omega_{b}$ appears unstable and oscillates between $\omega_{2}^{I} = \Omega/2 + \Omega^{\prime}$ and $\omega_{3}^{I} = -\Omega/2 + \Omega^{\prime}$ when $E < -1/2$, where the imaginary part $\Omega^{\prime}$ is stable but the real part oscillates between $\pm\Omega/2$, and $\omega_{2}^{IV} = -i\,\Omega/2 + |\Omega^{\prime}|$ and $\omega_{1}^{IV} = 
i\,\Omega/2 + |\Omega^{\prime}|$ when $E > 1/2$, where the real part $|\Omega^{\prime}|$ is stable but the imaginary part oscillates between 
$\pm i\,\Omega/2$. 

The source of this unstable behavior for $\omega_{b}$ in regions (I) and (IV), where $|E| > 1/2$, is unclear so Table \ref{tab:periods_Mathematica} can be used for a consistent calculation of the half-periods in regions (I) and (IV). We note, however, that this instability does not affect the identity 
$x_{2}^{I} = x_{3}^{I*}$ in region (I) and the identity $x_{2}^{IV} = x_{1}^{IV*}$ in region (IV).

\begin{table}
\caption{\label{tab:periods_Mathematica}{\sf WeierstrassHalfPeriods}$(g_{2},g_{3}) = \{\omega_{a},\omega_{b}\}$ as calculated by 
{\sf Mathematica}.} 
\begin{ruledtabular} 
\begin{tabular}{cccc}
 & $(g_{3},\Delta)$  & $\omega_{a}$                                    &  $\omega_{b}$                                 \\ \hline 
(I) & $(+,-)$                & $\omega_{1} = \Omega$                    &  $\omega_{2} = \Omega/2 + \Omega^{\prime}$   \\
 & $(+,0)$               & $\Omega_{0}$                                    & $i\infty$ \\
(II) & $(+,+)$               & $\omega_{1} = \omega$                    &  $\omega_{3} = \omega^{\prime}$                          \\ 
 & $(0,+)$                & $\omega_{0}$                                    & $i\,\omega_{0}$ \\
(III) & $(-,+)$                & $\omega_{3} = -i\,\omega$                  &  $\omega_{1} = -i\,\omega^{\prime} = |\omega^{\prime}|$                                      \\
 & $(-,0)$                & $-i\,\Omega_{0}$                                & $\infty$ \\
(IV) & $(-,-)$                & $\omega_{3} = -i\,\Omega$                 &  $\omega_{2} = |\Omega^{\prime}| - i\,\Omega/2$                                    
\end{tabular}
\end{ruledtabular}
\end{table}

\section{\label{sec:W_orbits}Weierstrass Orbit Solutions}

We now express the orbit solutions \eqref{eq:x_wp} for each of the four energy-level regions shown in Table \ref{tab:energy_levels}. These orbit solutions are shown in Fig.~\ref{fig:Weierstrass_cubic_orbits}.

\subsection{Orbits in Region (I)}

For region (I), with energy levels $E < -1/2$ and $(g_{3},\Delta) = (+,-)$, only unbounded (u) motion is possible since $x_{1}^{I} > 1$ is real and 
$x_{2}^{I} = x_{3}^{I*}$ are complex-conjugate. Using the initial condition $x_{I}^{u}(0) = x_{1}^{I}$ and $\omega_{1}^{I} = \Omega$ from Table \ref{tab:periods}, we find
\begin{equation}
x_{I}^{u}(t) \;=\; \wp\left(t + \omega_{1}^{I};\frac{}{} g_{2}, g_{3}\right) \;=\; \wp\left(t + \Omega;\frac{}{} g_{2}, g_{3}\right).
\label{eq:xI_u} 
\end{equation}
Since $\Omega$ is real and $\wp(2\Omega; g_{2},g_{3}) = \wp(0;g_{2},g_{3}) = \infty$, then the complete unbounded orbit $x_{I}^{u}(t)$ can be mapped out for $-\Omega < t < \Omega$.

\subsection{Orbits in Region (II)}

For region (II), with energy levels $-1/2 < E < 0$ and $(g_{3},\Delta) = (+,+)$, unbounded (u) and bounded (b) motions are possible for $x_{II}^{u} > x_{1}^{II}$ and between $x_{3}^{II} < x_{II}^{b} < x_{2}^{II} < 0$. 

For the bounded-motion solution, using the initial condition $x_{II}^{b}(0) = x_{3}^{II}$ and $\omega_{3}^{II} = \omega^{\prime}$ from Table \ref{tab:periods}, we find the periodic solution
\begin{eqnarray}
x_{II}^{b}(t) & = & \wp\left(t + \omega_{3}^{II};\frac{}{} g_{2}, g_{3}\right) \nonumber \\
 & = & \wp\left(t + \omega^{\prime};\frac{}{} g_{2}, g_{3}\right),
\label{eq:xII_b} 
\end{eqnarray}
with a period $2\,\omega_{1}^{II} = 2\,\omega$, so the bounded orbit $x_{II}^{b}(t)$ can be mapped out for $0 \leq t \leq 2\,\omega$. At the 
half-period $t = \omega_{1}^{II} = \omega$, we find
\begin{eqnarray} 
x_{II}^{b}(\omega_{1}^{II}) & = & \wp\left(\omega_{1}^{II} + \omega_{3}^{II};\frac{}{} g_{2}, g_{3}\right) \nonumber \\
 & = & \wp\left(\omega_{2}^{II};\frac{}{} g_{2}, g_{3}\right) \;=\; x_{2}^{II}. 
 \end{eqnarray}
 We note that as the energy level approaches $E = -1/2$ (from above), the period approaches $2\Omega_{0} = 2\pi/\sqrt{6}$, which is the expected period of small oscillations in the vicinity of the stable equilibrium point at $x = -1/2$, where $V^{\prime\prime}(-1/2) = 6$. The bounded-motion solution \eqref{eq:xII_b} approaches $x_{II}^{b}(t) \rightarrow -1/2$ as $\omega^{\prime} \rightarrow i\,\infty$. 
 
For the unbounded-motion solution, using the initial condition $x_{II}^{u}(0) = x_{1}^{II}$ and $\omega_{1}^{II} = \omega$ from Table \ref{tab:periods}, we find
\begin{equation}
x_{II}^{u}(t) \;=\; \wp\left(t + \omega_{1}^{II};\frac{}{} g_{2}, g_{3}\right) \;=\; \wp\left(t + \omega;\frac{}{} g_{2}, g_{3}\right).
\label{eq:xII_u} 
\end{equation}
Once again, since $\omega$ is real and $\wp(2\omega; g_{2},g_{3}) = \wp(0;g_{2},g_{3}) = \infty$, then the complete unbounded orbit $x_{II}^{u}(t)$ can be mapped out for $-\omega < t < \omega$.

\subsection{Orbits in Region (III)}

For region (III), with energy levels $0 < E < 1/2$ and $(g_{3},\Delta) = (-,+)$, unbounded (u) and bounded (b) motions are possible for $x_{III}^{u} > x_{1}^{III}$ and between $x_{3}^{III} < x_{III}^{b} < x_{2}^{III}$, where $x_{3}^{III} < 0$ and $x_{2}^{III} > 0$. 

For the bounded-motion solution, using the initial condition $x_{III}^{b}(0) = x_{3}^{III}$ and $\omega_{3}^{III} = -i\,\omega$ from Table \ref{tab:periods}, we find the periodic solution
\begin{eqnarray}
x_{III}^{b}(t) & = & \wp\left(t + \omega_{3}^{III};\frac{}{} g_{2}, g_{3}\right) \;=\; \wp\left(t - i\,\omega;\frac{}{} g_{2}, g_{3}\right) \nonumber \\
 & = & -\;\wp\left(i\,t + \omega;\frac{}{} g_{2}, |g_{3}| \right),
\label{eq:xIII_b} 
\end{eqnarray}
where we used the inversion formula \eqref{eq:g3_inv}. This bounded solution has the period $2\,\omega_{1}^{III} = 2\,|\omega^{\prime}|$, so the bounded orbit $x_{III}^{b}(t)$ can be mapped out for $0 \leq t \leq 2\,|\omega^{\prime}|$. At at the half-period $t = \omega_{1}^{III} = |\omega^{\prime}|$, we find
\begin{eqnarray} 
x_{III}^{b}(\omega_{1}^{III}) & = & \wp\left(\omega_{1}^{III} + \omega_{3}^{III};\frac{}{} g_{2}, g_{3}\right) \nonumber \\
 & = & \wp\left(\omega_{2}^{III};\frac{}{} g_{2}, g_{3}\right) \;=\; x_{2}^{III}. 
 \end{eqnarray}
 We note that as the energy level approaches $E = 1/2$ (from below), the period approaches $2|\omega^{\prime}| \rightarrow \infty$ while the bounded-motion solution \eqref{eq:xIII_b} approaches the separatrix solution (with $g_{3} = -1$ and $\Delta = 0$)
\begin{eqnarray}
x_{III}^{bs}(t) & = & \wp\left(t - i\,\Omega_{0};\frac{}{} 3, -1\right) \nonumber \\
 & = & -1 + \frac{3}{2} \coth^{2}\left(\sqrt{\frac{3}{2}}\,(t - i\,\Omega_{0})\right),
 \label{eq:bounded_separatrix}
\end{eqnarray} 
where $\Omega_{0} = \pi/\sqrt{6}$. Here, we recover 
\[ x_{III}^{bs}(0) \;=\; -1 + \frac{3}{2} \coth^{2}\left(-i\,\frac{\pi}{2}\right) = -1 \]
and, as $t \rightarrow \infty$, we find $x_{III}^{bs}(t) \rightarrow -1 + 3/2 = 1/2$.

For the unbounded-motion solution, using the initial condition $x_{III}^{u}(0) = x_{1}^{III}$ and $\omega_{1}^{III} = -i\,\omega^{\prime} = 
|\omega^{\prime}|$ from Table \ref{tab:periods}, we find
 \begin{eqnarray}
x_{III}^{u}(t) & = & \wp\left(t + \omega_{1}^{III};\frac{}{} g_{2}, g_{3}\right) \;=\; \wp\left(t + |\omega^{\prime}|;\frac{}{} g_{2}, g_{3}\right) \nonumber \\
 & = & -\;\wp\left(i\,t + \omega^{\prime};\frac{}{} g_{2}, |g_{3}| \right),
\label{eq:xIII_u} 
\end{eqnarray}
where we used the inversion formula \eqref{eq:g3_inv}. Once again, since $|\omega^{\prime}|$ is real and $\wp(2\omega^{\prime}; g_{2},|g_{3}|) = \wp(0;g_{2},|g_{3}|) = \infty$, then the complete unbounded orbit $x_{III}^{u}(t)$ can be mapped out for $-|\omega^{\prime}| < t < |\omega^{\prime}|$. 

The unbounded separatrix solution $x_{III}^{bu}(t)$ may be obtained from a slight modification of the unbounded solution \eqref{eq:xIII_u}. First, we define the new unbounded solution in region (III) to be
 \begin{eqnarray}
x_{III}^{u}(t) & = & \wp\left(t;\frac{}{} g_{2}, g_{3}\right) \;=\; -\;\wp\left(i\,t;\frac{}{} g_{2}, |g_{3}| \right),
\label{eq:xIII_u_new} 
\end{eqnarray}
which assumes that $x_{III}^{u}(0) = \infty$ and 
\[ x_{III}^{u}(|\omega^{\prime}|) \;=\; \wp\left(|\omega^{\prime}|;\frac{}{} g_{2}, g_{3}\right) = x_{1}^{III}. \]
The unbounded separatrix solution may be written as
\begin{equation}
x_{III}^{us}(t) \;=\; -1 + \frac{3}{2} \coth^{2}\left(\sqrt{\frac{3}{2}}\,t\right),
 \label{eq:unbounded_separatrix}
\end{equation} 
where $x_{III}^{us}(0) = \infty$ and $x_{III}^{us}(\infty) = -1 + 3/2 = 1/2$.

\subsection{Orbits in Region (IV)}

Lastly, for region (IV), with energy levels $E > 1/2$ and $(g_{3},\Delta) = (-,-)$, only unbounded (u) motion is possible since $x_{3}^{IV} < -1$ is real and $x_{1}^{IV} = x_{2}^{IV*}$ are complex-conjugate. Using the initial condition $x_{IV}^{u}(0) = x_{3}^{IV} = -\,x_{1}^{I}$ and $\omega_{3}^{IV} = 
-i\,\Omega$ from Table \ref{tab:periods}, we find
\begin{eqnarray}
x_{IV}^{u}(t) & = & \wp\left(t + \omega_{3}^{IV};\frac{}{} g_{2}, g_{3}\right) \;=\; \wp\left(t - i\,\Omega;\frac{}{} g_{2}, g_{3}\right) \nonumber \\
 & = & -\;\wp\left(\Omega + i\,t;\frac{}{} g_{2}, |g_{3}|\right),
\label{eq:xIV_u} 
\end{eqnarray}
where we used the inversion formula \eqref{eq:g3_inv}. We note that at $t = 2\,|\Omega^{\prime}| \equiv \omega_{1}^{IV} + \omega_{2}^{IV}$, or
$i\,t = 2\,\Omega^{\prime}$, we find
\begin{eqnarray} 
x_{IV}^{u}(2\,|\Omega^{\prime}|) & = & -\;\wp\left(\Omega + 2\,\Omega^{\prime};\frac{}{} g_{2}, |g_{3}|\right)  \\
 & = & -\;\wp\left(2\,\omega_{2}^{I};\frac{}{} g_{2}, |g_{3}|\right) = -\;(-\,\infty) = \infty, \nonumber
 \end{eqnarray}
so that the complete unbounded orbit $x_{IV}^{u}(t)$ can be mapped out for $-2\,|\Omega^{\prime}| < t < 2\,|\Omega^{\prime}|$.

\subsection{Weierstrass orbit solutions}

\begin{figure}
\epsfysize=2.5in
\epsfbox{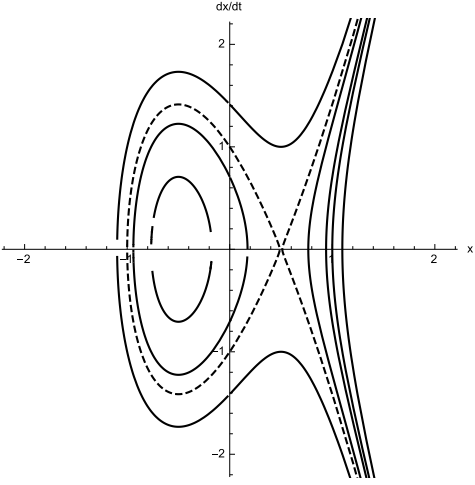}
\caption{Phase plot $(x,dx/dt)$ for the Weierstrass unbounded and bounded orbit solutions. The bounded and unbounded separatrix solutions are shown as dashed curves and the square dot denotes the stable equilibrium point at $x = -1/2$.}
\label{fig:Weierstrass_cubic_orbits}
\end{figure}

The Weierstrass orbit solutions are shown in Fig.~\ref{fig:Weierstrass_cubic_orbits}, with the bounded and unbounded separatrix solutions (for 
$g_{3} = -1$ and $\Delta = 0$)  shown as dashed curves. Solutions found inside the bounded separatrix solution are the bounded solutions \eqref{eq:xII_b} and \eqref{eq:xIII_b}, while solutions found outside are the unbounded solutions \eqref{eq:xI_u}, \eqref{eq:xII_u}, \eqref{eq:xIII_u}, and \eqref{eq:xIV_u}. The bounded solution \eqref{eq:xII_b}  in region (II) is periodic between $x_{3}^{II} < x_{2}^{II} < 0$, while the bounded solution \eqref{eq:xIII_b} in region (III) is periodic between $x_{3}^{III} < 0 <  x_{2}^{III}$. The unbounded solutions [\eqref{eq:xI_u}, \eqref{eq:xII_u}, \eqref{eq:xIII_u}] in regions (I)-(III) have a turning points at $0 < x_{1}^{III} < x_{1}^{II} < x_{1}^{I}$, while the unbounded solution \eqref{eq:xIV_u} in region (IV) has a turning point at $x_{3}^{IV} < 0$. Lastly, the potential \eqref{eq:V_def} has stable equilibrium point at $x = -1/2$ (shown as a square dot in Fig.~\ref{fig:Weierstrass_cubic_orbits}) and an unstable equilibrium point $x = 1/2$, shown in Fig.~\ref{fig:Weierstrass_cubic_orbits} as the point where the bounded and unbounded separatrix solutions cross.

Lastly, by using the $g_{3}$-inversion formula \eqref{eq:g3_inv}, we observe the following relations between bounded and unbounded orbit solutions
\begin{equation}
\left. \begin{array}{rcl}
x_{II}^{b}(t) & = & -\;x_{III}^{u}(i\,t)  \\
x_{III}^{b}(t) & = & -\;x_{II}^{u}(i\,t) \\
x_{IV}^{u}(t) & = & -\;x_{I}^{u}(i\,t)
\end{array} \right\}.
\label{eq:ub_relations}
\end{equation} 
Hence, bounded motion can be viewed in terms of unbounded motion in imaginary time.

\section{\label{sec:J_orbits}Jacobi Orbit Solutions}

In this last Section, we proceed with the transformation of our Weierstrass orbit solutions into expressions involving the Jacobi elliptic functions
\cite{NIST_Jacobi} ${\rm cn}(z|m)$, ${\rm sn}(z|m)$, and ${\rm dn}(z|m)$, where $z(t;\phi) = \kappa(\phi)\,t$ and the modulus $m(\phi)$ is defined as
\begin{eqnarray}
m(\phi) & \equiv & \frac{x_{2}(\phi) - x_{3}(\phi)}{x_{1}(\phi) - x_{3}(\phi)} \;=\;\frac{\sin(\phi/3)}{\sin[(\pi + \phi)/3]}.
\label{eq:m_phi_def}
\end{eqnarray}
Here, for $-1/2 < E < 0$ (region II), we have $0 < \phi < \pi/2$ and $0 < m < 1/2$, while for $0 < E < 1/2$ (region III), we have $\pi/2 < \phi < \pi$ and
$1/2 < m < 1$. For $|E| > 1/2$ (regions I and IV),  $m$ is a complex-valued number:
\begin{equation}
m(\psi) \;=\; \left\{ \begin{array}{lcr}
1 \;-\; \exp\left[i\frac{}{}\chi(\psi)\right] &  & (E < -1/2) \\
\exp\left[i\frac{}{}\chi(\psi)\right] &  & (E > 1/2)
 \end{array} \right. 
 \label{eq:m_psi}
 \end{equation}
 where the phase $\chi(\psi) = 2\;{\rm arctan}[3^{-\frac{1}{2}}\tanh(\psi/3)]$ vanishes at $\psi = 0$ and reaches $\chi(\infty) = 2\,{\rm arctan}(3^{-1/2}) = \pi/3$ at $\psi = \infty$. Hence, the relation between $m_{I}$ (for $E < -1/2$) and $m_{IV}$ (for $E > 1/2$) is $m_{I}^{\prime} \equiv 1 - m_{I} = m_{IV}$, where
\begin{eqnarray}
m^{\prime}(\phi) & \equiv & \frac{x_{1}(\phi) - x_{2}(\phi)}{x_{1}(\phi) - x_{3}(\phi)} \;=\;\frac{\sin[(\pi - \phi)/3]}{\sin[(\pi + \phi)/3]}.
\label{eq:mprime_phi_def}
\end{eqnarray}
For energy levels $|E| > 1/2$, we easily verify that $m_{IV}^{\prime}(\psi) \equiv m_{I}^{\prime}(\psi)$.

\subsection{Jacobi elliptic functions}

The Jacobi elliptic functions ${\rm cn}(z|m)$, ${\rm sn}(z|m)$, and ${\rm dn}(z|m)$ have real and imaginary periods determined by the complete elliptic integrals of the first kind ${\sf K}(m)$ and ${\sf K}^{\prime}(m) \equiv {\sf K}(m^{\prime}) = {\sf K}(1 - m)$, respectively:
\begin{equation}
\left. \begin{array}{rcl} 
{\rm cn}\left(z + 4n\,{\sf K} + 2n^{\prime}\,({\sf K} + i\,{\sf K}^{\prime})\frac{}{}|m\right) & = & {\sf cn}(z|m) \\
{\rm sn}\left(z + 4n\,{\sf K} + 2n^{\prime}\, i\,{\sf K}^{\prime}\frac{}{}|m\right) & = & {\sf sn}(z|m) \\
{\rm dn}\left(z + 2n\,{\sf K} + 4n^{\prime}\,i\,{\sf K}^{\prime}\frac{}{}|m\right) & = & {\sf dn}(z|m) 
\end{array} \right\},
\end{equation}
for all integers $n$ and $n^{\prime}$. In addition, these functions have zeroes at ${\rm cn}({\sf K}|m) = 0$, ${\rm sn}(0|m) = 0$, and 
${\rm dn}({\sf K} + i\,{\sf K}^{\prime}|m) = 0$, as well as singularities (poles) at $z = i\,{\sf K}^{\prime}$. Lastly, they satisfy the following identities ${\rm cn}^{2}(z|m) + 
{\rm sn}^{2}(z|m) = 1$,  ${\rm dn}^{2}(z|m) + m\;{\rm sn}^{2}(z|m) = 1$, and have the following limits: ${\rm sn}(z|0) = \sin z$, ${\rm cn}(z|0) = \cos z$, ${\rm dn}(z|0) = 1$, and 
${\rm sn}(z|1) = \tanh z$, ${\rm cn}(z|1) = {\rm sech}\,z = {\rm dn}(z|1).$ 

The function $y = {\rm sn}(z|m)$ satisfies the differential equation
\begin{equation}
\left(y^{\prime}\right)^{2} \;=\; \left(1 \;-\frac{}{} y^{2}\right) \;\left(1 \;-\frac{}{} m\;y^{2}\right),
\label{eq:sn_diff}
\end{equation}
the function $y = {\rm cn}(z|m)$ satisfies the differential equation
\begin{equation}
\left(y^{\prime}\right)^{2} \;=\; \left(1 \;-\frac{}{} y^{2}\right) \;\left(m^{\prime} \;+\frac{}{} m\;y^{2}\right),
\label{eq:cn_diff}
\end{equation}
and the function $y = {\rm dn}(z|m)$ satisfies the differential equation
\begin{equation}
\left(y^{\prime}\right)^{2} \;=\; \left(1 \;-\frac{}{} y^{2}\right) \;\left(y^{2} \;-\frac{}{} m^{\prime}\right).
\label{eq:dn_diff}
\end{equation}
Other Jacobi elliptic functions that will be useful below are $y = {\rm sc}(z|m) \equiv {\rm sn}(z|m)/{\rm cn}(z|m)$, which satisfies the differential equation
\begin{equation}
\left(y^{\prime}\right)^{2} \;=\; \left(1 \;+\frac{}{} y^{2}\right) \;\left(1 \;+\frac{}{} m^{\prime}\;y^{2} \right),
\label{eq:sc_diff}
\end{equation}
and $y = {\rm cs}(z|m) \equiv {\rm cn}(z|m)/{\rm sn}(z|m)$, which satisfies the differential equation
\begin{equation}
\left(y^{\prime}\right)^{2} \;=\; \left(1 \;+\frac{}{} y^{2}\right) \;\left(m^{\prime} \;+\frac{}{} y^{2} \right).
\label{eq:cs_diff}
\end{equation}
By comparing Eqs.~\eqref{eq:sn_diff} and \eqref{eq:sc_diff}, for example, we obtain the following identity
\begin{equation}
{\sf sn}(z|m^{\prime}) \;=\; -\,i\;{\rm sc}(i z|m),
\label{eq:sn_sc}
\end{equation}
while Eq.~\eqref{eq:cn_diff} yields the identity ${\rm cn}(z|m^{\prime}) = 1/{\rm cn}(iz|m) \equiv {\rm nc}(iz|m) $.

\subsection{Bounded orbit solutions}

We begin with bounded orbit solutions in region (II) and (III), where $0 \leq m(\phi) \leq 1$. First, we substitute 
\begin{equation}
x^{b}(t) \;=\; x_{3} + \alpha\,y^{2}(\kappa t) 
\label{eq:b_Jacobi}
\end{equation}
into Eq.~\eqref{eq:W_form_roots} to obtain
\[ \alpha\kappa^{2}\,\left(y^{\prime}\right)^{2} \;=\; \left[(x_{1} - x_{3}) - \alpha y^{2}\right]\;\left[(x_{2} - x_{3}) - \alpha y^{2}\right]. \]
Next, using the definition \eqref{eq:m_phi_def}, we define $\kappa^{2} = x_{1} - x_{3}$ and $\alpha = x_{2} - x_{3} \equiv m\,\kappa^{2}$, so that we obtain the differential equation \eqref{eq:sn_diff}, whose solution is $y(\kappa t) = {\sf sn}(\kappa t\,|\,m)$. Hence, in regions (II) and (III), the bounded orbit solution is
\begin{equation}
x^{b}(t) \;=\; x_{3} + (x_{2} - x_{3})\;{\sf sn}^{2}\left(\sqrt{x_{1} - x_{3}}\,t\;\left|\frac{}{}\right. m\right).
\label{eq:xb_sn}
\end{equation}
The bounded separatrix solution is obtained from Eq.~\eqref{eq:xb_sn} in the limit $m \rightarrow 1$:
\begin{equation}
x^{bs}(t) \;=\; -1 + \frac{3}{2}\;\tanh^{2}\left(\sqrt{\frac{3}{2}}\,t\right),
\label{eq:xb_sep}
\end{equation}
which is identical to Eq.~\eqref{eq:bounded_separatrix}.

Since the Jacobi elliptic function ${\sf sn}(z|m)$ has a period $4{\sf K}(m)$, where ${\sf K}(m)$ is the complete elliptic integral of the first kind, the period of the orbit solution \eqref{eq:xb_sn} is 
\begin{equation}
\frac{2\,{\sf K}(m)}{\sqrt{x_{1} - x_{3}}} \;=\; \left\{ \begin{array}{lcr}
2\,\omega & & (0 \leq m \leq 1/2) \\
 & & \\
2\,|\omega^{\prime}| & & (1/2 \leq m < 1)
\end{array} \right.
\label{eq:Jacobi_b}
\end{equation} 
For $E = -1/2$, we find $m = 0$, with $x_{1} - x_{3} = 3/2$ and $x_{2} - x_{3} = 0$, so that the period \eqref{eq:Jacobi_b} is $2{\sf K}(0)/\sqrt{3/2} = 
2\pi/\sqrt{6} = 2\Omega_{0}$, where we used ${\sf K}(0) = \pi/2$. In the limit $m \rightarrow 1$ $(E \rightarrow 1/2)$, $|\omega^{\prime}| \rightarrow 
\infty$ since ${\sf K}(m) \rightarrow {\sf K}(1) = {\sf K}^{\prime}(0) = \infty$. When $E = 0$ ($m = 1/2$), we find $x_{1} - x_{3} = \sqrt{3}$ and the period \eqref{eq:Jacobi_b} is 
$2{\sf K}(1/2)/3^{1/4} = 2\omega_{0}$.

\subsection{Unbounded orbit solutions}

\subsubsection{Regions (I)-(III)}

We now consider the unbounded orbit solutions in regions (I)-(III), where $x^{u}(0) = x_{1}$. First, we substitute 
\begin{equation}
x^{u}(t) \;=\; x_{1} \;+\; \alpha/y^{2}(\kappa t) 
\label{eq:u_Jacobi}
\end{equation}
into Eq.~\eqref{eq:W_form_roots}  to obtain
\[ \alpha\kappa^{2}\left(y^{\prime}\right)^{2} \;=\; \left[y^{2}(x_{1} - x_{2}) + \alpha\right]\;\left[y^{2}(x_{1} - x_{3}) + \alpha\right], \]
where the solution $y(\kappa t)$ must satisfy the initial condition $y(0) = \infty$. Next, using the definition \eqref{eq:mprime_phi_def}, we define $\kappa^{2} = x_{1} - x_{3}$ 
and $\alpha = x_{1} - x_{2} = m^{\prime}\,\kappa^{2}$, so that we obtain the differential equation \eqref{eq:cs_diff}, whose solution is  $y(\kappa t) = {\sf cn}(\kappa t|m)/{\sf sn}(\kappa t|m)$. Hence, in regions (I)-(III), the unbounded orbit solution is
\begin{equation}
x^{u}(t) \;=\; x_{1} \;+\; (x_{1} - x_{2})\;\frac{{\sf sn}^{2}(\sqrt{x_{1} - x_{3}}\,t\;|\;m)}{{\sf cn}^{2}(\sqrt{x_{1} - x_{3}}\,t\;|\;m)}.
\label{eq:xu_sc}
\end{equation}
In regions (II) and (III), all roots $x_{3} < x_{2} < x_{1}$ are real and $0 < m < 1$. The unbounded orbit solution \eqref{eq:xu_sc} reaches infinity after a time ${\sf K}(m)/
\sqrt{x_{1} - x_{3}}$ when ${\sf cn}({\sf K}(m)|m) = 0$. In region (I), where $x_{1}$ is real and $x_{2} = x_{3}^{*}$ are complex-conjugate roots, and the definition \eqref{eq:m_psi} for $m$ implies that it is complex-valued.

We note that the unbounded separatrix solution cannot be obtained from Eq.~\eqref{eq:xu_sc} since $x_{1} = x_{2}$ when $E = 1/2$ (or $m = 1$). If we use the initial condition $x^{bu}(0) = \infty$, i.e., $y(0) = 0$, and the definitions $\alpha = x_{1} - x_{3} = \kappa^{2}$, we obtain the differential equation \eqref{eq:sc_diff}, whose solution is $y(\kappa t) = {\sf sn}(\kappa t|m)/{\sf cn}(\kappa t|m)$. Hence, the unbounded orbit solution in region (III) can also be expressed as
\begin{equation}
x^{u}(t) \;=\; x_{1} \;+\; (x_{1} - x_{3})\;\frac{{\sf cn}^{2}(\sqrt{x_{1} - x_{3}}\,t\;|\;m)}{{\sf sn}^{2}(\sqrt{x_{1} - x_{3}}\,t\;|\;m)},
\label{eq:xu_cs}
\end{equation}
so that the unbounded separatrix solution becomes
\begin{eqnarray}
x^{us}(t) & = & \frac{1}{2} \;+\; \frac{3}{2}\;\frac{{\rm sech}^{2}(\sqrt{3/2}\,t)}{{\rm tanh}^{2}(\sqrt{3/2}\,t)} \nonumber \\
 & = & -1 \;+\; \frac{3}{2}\coth^{2}\left(\sqrt{\frac{3}{2}}\,t\right),
\label{eq:xus_cs}
\end{eqnarray}
which is identical to Eq.~\eqref{eq:unbounded_separatrix}.

\subsubsection{Region (IV)}

In region (IV), where $E > 1/2$, $x_{3} < 0$, and $x_{1} = x_{2}^{*}$, the parameter $m$ is a complex-valued number \eqref{eq:m_psi} with unit magnitude $|m| = 1$. The unbounded orbit solution
\begin{equation}
x_{IV}^{u}(t) \;=\; x_{3} + (x_{2} - x_{3})\;{\sf sn}^{2}\left(\sqrt{x_{1} - x_{3}}\,t\;\left|\frac{}{}\right. m\right)
\label{eq:xu_sn}
\end{equation}
given by Eq.~\eqref{eq:xb_sn} is still applicable in this region. Using the fact that ${\sf sn}(z|m)$ is infinite at $z = i\,{\sf K}^{\prime}(m)$ for $m \neq 0$. Hence, at a finite (complex) time $i\,{\sf K}^{\prime}(m)/\sqrt{x_{1} - x_{3}}$, the particle reaches infinity. 

\subsection{Relations between bounded and unbounded orbit solutions}

We now wish to verify that the Jacobi orbit solutions derived in this Section satisfy the relations \eqref{eq:ub_relations} between the bounded and unbounded solutions.

If we use the relation \eqref{eq:sn_sc}, Eq.~\eqref{eq:xu_sn} becomes
\begin{eqnarray}
x_{IV}^{u}(t) & = & x_{3}^{IV} + \alpha_{IV}\;{\sf sn}^{2}\left(\kappa_{IV}\,t\;| m_{IV}\right) 
\label{eq:xu_sn_mprime} \\
 & = & x_{3}^{IV} - \alpha_{IV}\;\frac{{\sf sn}^{2}\left(i\,\kappa_{IV}\,t\;| m_{I}\right)}{
 {\sf cn}^{2}\left(i\,\kappa_{IV}\,t\;|m_{I}\right)} \nonumber \\
  & st= & -\,x_{1}^{I} \;-\; \alpha_{I}\;\frac{{\sf sn}^{2}\left(i\,\kappa_{I}\,t\;|m_{I}\right)}{
 {\sf cn}^{2}\left(i\,\kappa_{I}\,t\;| m_{I}\right)} \;\equiv\; -\;x_{I}^{u}(i\,t),
\nonumber
\end{eqnarray}
where we used the fact that $m_{IV} = m_{I}^{\prime} = 1 - m_{I}$ and we recover the relation \eqref{eq:ub_relations}.  Next, we look at the bounded solution in region (III)
\begin{eqnarray}
x_{III}^{b}(t) & = & x_{3}^{III} + \alpha_{III}\;{\sf sn}^{2}\left(\kappa_{III}\,t\;|m_{III}\right)  \label{eq:xb_sn_mprime} \\
  & = & -\,x_{1}^{II} - \alpha_{II}\;\frac{{\sf sn}^{2}\left(i\,\kappa_{II}\,t\;|m_{II}\right)}{
 {\sf cn}^{2}\left(i\,\kappa_{II}\,t\;| m_{II}\right)} \;=\; -\,x_{II}^{u}(i\,t),
\nonumber
\end{eqnarray}
and the bounded solution in region (II)
\begin{eqnarray}
x_{II}^{b}(t) & = & x_{3}^{II} + \alpha_{II}\;{\sf sn}^{2}\left(\kappa_{II}\,t\;|m_{II}\right) \nonumber \\
  & = & -\,x_{1}^{III} - \alpha_{III}\;\frac{{\sf sn}^{2}\left(i\,\kappa_{III}\,t\;|m_{III}\right)}{
 {\sf cn}^{2}\left(i\,\kappa_{III}\,|m_{III}\right)} \nonumber \\
  & = & -\,x_{III}^{u}(i\,t).
 \label{eq:xb_sn_m}
\end{eqnarray}
Hence, the relations \eqref{eq:ub_relations}, which were originally obtained through the $g_{3}$-inversion relation \eqref{eq:g3_inv}, are also obtained through the transformation $m \rightarrow m^{\prime} = 1 - m$, even when $m$ is complex-valued.

\section{Summary}

A consistent notation for the half-periods $(\omega_{1},\omega_{3},\omega_{2} \equiv \omega_{1} + \omega_{3})$ of the Weierstrass elliptic function $\wp(z; g_{2},g_{3})$ was constructed from the analysis of the motion of a particle in a cubic potential. A guide to the accurate interpretation of the 
{\sf Mathematica} output for the function {\sf WeierstrassHalfPeriods}$(g_{2},g_{3}) = \{\omega_{a},\omega_{b}\}$ was also provided.


\begin{thebibliography}{99}

\bibitem{NIST_Weierstrass} W.~P.~Reinhardt and P.~L.~Walker, {\it Weierstrass Elliptic and Modular Functions}, in NIST Handbook of Mathematical Functions (Cambridge University Press, Cambridge, 2010), Chap.~23.

\bibitem{Lawden} D.~F.~Lawden, {\it Elliptic Functions and Applications}, (Springer-Verlag, New York, 1989).

\bibitem{Brizard_2007} A.~J.~Brizard,  {\it A primer on elliptic functions with applications in classical mechanics}, arXiv:0711.4064v1 (2007).

\bibitem{Brizard_2009} A.~J.~Brizard, Eur.~J.~Phys.~{\bf 30}, 729 (2009).

\bibitem{Brizard_Lag} A.~J.~Brizard, {\it An Introduction to Lagrangian Mechanics}, 2nd ed. (World Scientific, 2015), App.~B.

\bibitem{NIST_Jacobi} W.~P.~Reinhardt and P.~L.~Walker, {\it Jacobian Elliptic Functions}, in NIST Handbook of Mathematical Functions (Cambridge University Press, Cambridge, 2010), Chap.~22.


\end{thebibliography}
\end{document}